# Temperature Dependent Raman Studies and Thermal Conductivity of Few Layer MoS$_2$

Satyaprakash Sahoo,* Anand P.S. Gaur, Majid Ahmadi, Maxime J-F Guinel and Ram S. Katiyar*

Department of Physics and Institute for Functional Nanomaterials, University of Puerto Rico, San Juan, PR 00931 USA

We report on the temperature dependence of in-plane $E_{2g}$ and out of plane $A_{1g}$ Raman modes in high quality few layers MoS$_2$ (FLMS) prepared using a high temperature vapor-phase method. The materials obtained were investigated using transmission electron microscopy. The frequencies of these two phonon modes were found to vary linearly with temperature. The first order temperature coefficients for $E_{2g}$ and $A_{1g}$ modes were found to be $1.32 \times 10^{-2}$ and $1.23 \times 10^{-2}$ cm$^{-1}$/K, respectively. The thermal conductivity of the suspended FLMS at room temperature was estimated to be about 52 W/mK.

*Corresponding authors:

Email: satya504@gmail.com (S. Sahoo), rkatiyar@hpcf.upr.edu (R. S. Katiyar)

Atomically thin two dimensional (2D) materials such as graphene and hexagonal boron nitride (h-BN) have attracted significant research interests because of their extraordinary physical properties.[1,2,3,4,5,6] However, graphene is a zero band gap material whose Dirac energy states meet with each other at the *K* point of the Brillouin zone.[7] The lack of band gap is undesirable for electronic applications, particularly in transistors. In contrast, h-BN is a high band gap material with a band gap of 5.2 eV.[8] On the other hand, the metal dichalcogenides ($MX_2$ where M=metal and X= S, Se or Te) share many common properties with graphene including the two dimensional layered structure, and can have a finite band gap.[9,10] These semiconductors are now considered as the possible alternatives to graphene and currently of great research interest. Among the metal dichalcogenides, $MoS_2$ is one of the most stable layered material and is an indirect band gap semiconductor (band gap about 1.3 eV). Its band gap increases when the number of layers is decreased reaching about 1.8 eV (direct band gap) for monolayer $MoS_2$.[11,12,13] In the 2D hexagonal lattice of a monolayer $MoS_2$, each Mo is six fold coordinated and sandwiched between two three coordinated S atoms in a trigonal prismatic arrangement. Both single layer and FLMS are attractive for low power electronic devices.[14,15,16] The charge mobility of $MoS_2$ can be altered by the gate material. Similar to graphene, $MoS_2$ has strong Raman scattering intensity.[17,18] The Raman peak positions depend upon the number of layers and thus it is possible to determine the number of layers from the Raman spectra.[18]

The knowledge of the vibrational properties of FLMS is important to understand the electron-phonon interaction and the transport properties which in turn have a large impact on the electronic device performances. Vibrational properties are also of fundamental importance to understand anharmonicity in the lattice potential energy. Secondly, the performance of electronic devices (size of a few micro/nanometers) largely depends on the nature of the heat dissipation and hence on the thermal conductivity. Although 2D hexagonal materials are of great importance for future nano-electronics, the determination of the thermal conductivity has been limited to single and bilayer layer graphene and a few layer h-BN.[19,20] The thermal conductivity in layered materials such as graphite is anisotropic in nature: the in-plane thermal conductivity is much higher than cross-plane.[21] Recently, ultra high thermal conductivity (4000 to 5000 W/mK) was measured in single layer graphene using Raman spectroscopy.[19] Raman scattering is an indirect method for evaluating the thermal conductivity for materials and its accuracy solely depends on the intensity of the Raman signal. In case of h-BN, the Raman intensity is weak and it further weakens when the number of layers is decreased. Hence, Raman scattering is not suitable to estimate the thermal conductivity of h-BN. Jo *et al.* have reported the thermal conductivity (230W/mK) of few layers h-BN using micro bridge resistive thermometer method.[20] In contrast to BN, $MoS_2$ has strong Raman signal and the intensity of Raman signal increases with decreases in layer thickness. Thus Raman spectroscopic studies on suspended $MoS_2$ can be helpful to estimate the thermal conductivity.

In this paper, we report on the temperature dependent Raman studies of high quality FLMS prepared using a vapor-phase method. The FLMS materials were also studied using transmission electron microscopy (TEM). Furthermore, we evaluate the thermal conductivity of a suspended FLMS by performing a laser power dependence Raman study of the suspended FLMS. To the best of our knowledge, there are no previous reports either on the temperature dependent Raman studies on FLMS or thermal conductivity value for bulk $MoS_2$.

The FLMS were prepared on a thick $SiO_2$ (300 nm)/Si substrate by a modified high temperature vapor-phase method.[22] Sulfurization of desired Mo films was carried out in a tube furnace in atmosphere containing argon and hydrogen gases in the ratio of 9:1 at 900°C. Bulk

MoS$_2$ was purchased from SPI. Raman measurements were using a Horiba-Yobin T64000 micro-Raman system, and a 532 nm wavelength radiation from a diode laser as an excitation source. The samples were characterized using a cold field emission scanning electron microscope (FESEM, JEOL JSM-7500F SEM), a TEM (Carl Zeiss LEO 922) and a high resolution TEM (HRTEM, JEOL JEM-2200FS). Etching the SiO$_2$ using a KOH solution produced suspended FLMS that were placed on a TEM copper (no carbon film) grids (**supplementary information Fig. S1**).

Figure 1(a) shows a TEM image recorded from a FLMS. The number of layers composing the FLMS was determined by examining the edges of the sheets parallel to the electron beam. An example is shown in Fig. 1(b) where the specific FLMS consisted of 11 layers and measuring about 6.6 nm thick with the distance between consecutive layers measured to about ~0.60 nm. This is in accordance with the (002) lattice spacing measured using XRD and confirmed with the JCPDS card No.73-1508. The distance between layers of MoS$_2$ can vary from 0.6 to 0.7 nm.[23] Figure 1(c, d) shows HRTEM images recorded from the FLMS material. The lattice spacing for (100) and (103) were measured to 0.27 and 0.23 nm, respectively, which are consistent with the[23] and JCPDS No. 73-1508.

During micro-Raman experiments it is important to use appropriate laser power as high laser powers can significantly increase the local temperature on the sample.[24,25] This often affects the Raman spectra by broadening and shifting of the Raman peaks in addition to the possibility of damaging the sample. We found that an increase in laser power up to 2 mW has no effect on the peak position and broadening of Raman peaks in SiO$_2$/Si supported FLMS. The crystal lattice of the layered bulk MoS$_2$ belongs to D$_{6h}$ point group and the four Raman active zone centre phonon modes can be represented by the following irreducible representation;

$\Gamma = A_{1g} + 2E_{2g} + E_{1g}$.[26] The room temperature Raman spectra recorded for FLMS on the SiO$_2$/Si substrate and exfoliated bulk MoS$_2$ are shown in Fig. 2(a). Both samples have characteristic peaks at about 32, 383 and 409 cm$^{-1}$ which are assigned to $E^2_{2g}$, $E^1_{2g}$ and $A_{1g}$ modes, respectively.[17,18] In a back scattering geometry, the $E_{1g}$ mode is forbidden for the plane perpendicular to the c-axis, i.e. the basal plane. This mode was neither observed for FLMS nor bulk MoS$_2$. The relative motion of the various atoms for these three observed phonons is schematically presented in Fig. 2(b). Note that $E^2_{2g}$ is the shear mode and is due to the relative motion between two monolayers. The frequency of vibrations of these phonon modes depend on the number of layers and a systematic change has been observed as the number of layers increases from 1 to 7.[27] The thickness dependent vibrational frequency of the $E^2_{2g}$ is mainly due to the weak Van der Waals interlayer interaction. In the present study the full width at half maximum of the Raman peaks ($E^1_{2g}$ and $A_{1g}$ modes) for exfoliated MoS$_2$ and FLMS are very close to each other. The FWHM of the $E^1_{2g}$ and $A_{1g}$ modes are found to be 2.7 and 3.1 cm$^{-1}$, respectively in both bulk and FLMS. This suggests our FLMS material has high crystallinity. The non-vanishing polarizability tensor for these Raman modes is given below,[28]

$$A_{1g} = \begin{bmatrix} a & 0 & 0 \\ 0 & a & 0 \\ 0 & 0 & b \end{bmatrix}, \quad E_{2g} = \begin{bmatrix} 0 & d & 0 \\ d & 0 & 0 \\ 0 & 0 & 0 \end{bmatrix}, \begin{bmatrix} 0 & d & 0 \\ 0 & -d & 0 \\ 0 & 0 & 0 \end{bmatrix}$$

The temperature dependent Raman studies on FLMS were carried out by varying the temperature starting from near liquid nitrogen temperature (about 83 K) to 523 K. Figure 3 (a) shows the temperature dependent Raman spectra of FLMS. For all temperatures there are strong scattering intensities from both $E^1_{2g}$ and $A_g$ modes. As seen from the figure that both $E^1_{2g}$ and $A_{1g}$ modes follow systematic red-shift with increase in temperature. The full width at half maximum (FWHM) of both peaks increases with increasing temperature. A discussion on temperature dependent boarding of $A_{1g}$ mode will be discussed latter. Note that, for single and bilayer graphene, the G peak frequency has been reported to decrease with increase in temperature however, the FWHM was unaffected by same range of temperature change.[29] The change in peak position of the $E^1_{2g}$ and $A_{1g}$ modes with increase in the temperature are plotted in Fig. 4(b) and (c), respectively. The data of the peak position versus temperature were fitted using following equation:

$$\omega(T)=\omega_o+\chi T \qquad (1)$$

where $\omega_o$ is the frequency of vibration of the $E^1_{2g}$ or $A_g$ modes at absolute zero temperature, $\chi$ is first order temperature coefficient of the $E^1_{2g}$ or $A_g$ modes. The slope of the fitted straight line represents the value of $\chi$. The values of $\chi$ for $E^1_{2g}$ and $A_g$ modes are found to be -1.32 $10^{-2}$ and -1.23×$10^{-2}$ cm$^{-1}$/K, respectively. It may be emphasized that we did not consider the higher order temperature coefficients as these terms are significant only at high temperatures. Interestingly, the frequency of $E^2_{2g}$ mode is not affected much by rise in temperature; a shift of ~ 0.6 cm$^{-1}$ in the peak position was noticed for the whole temperature range (83 to 523K). However, the intensity of $E^2_{2g}$ mode increases with increase in temperature (**supplementary information Fig. S2**).

The behavior of the change in Raman peak position with temperature varies for different materials, even for a given material the change in phonon frequency with temperature may differ for different phonon modes. For example, in anatase (TiO$_2$) the phonon frequencies of two of the $E_g$ modes blue-shift with increase in temperature where as one of the $E_g$ and $B_{1g}$ mode red-shift with increase in temperature.[30] The variation in the Raman peak position of the normal modes with temperature is mainly due to the contribution from thermal expansion or volume contribution and from temperature contribution which results from anharmonicity. The phonon frequency $\omega$ can be expressed as a function of volume and temperature as follows,[31]

$$\left(\frac{\partial ln\omega}{\partial T}\right)_P = \left(\frac{\partial lnV}{\partial T}\right)_P \left(\frac{\partial ln\omega}{\partial lnV}\right)_T + \left(\frac{\partial ln\omega}{\partial T}\right)_V = -\frac{\gamma}{k}\left(\frac{\partial ln\omega}{\partial P}\right)_T + \left(\frac{\partial ln\omega}{\partial T}\right)_V \qquad (2)$$

where $\gamma \sim (\partial lnV/\partial T)_P$ and $k \sim -(\partial lnV/\partial P)_T$ are the volume thermal coefficient and isothermal volume compressibility. The first term of the right hand side of the equation represents the volume contribution at constant temperature and the second term represents the temperature contribution at constant volume. Hence anharmonic (pure-temperature) contribution can be determined from the values of $\gamma$, $k$ and isobaric temperature and isothermal pressure derivative of phonon frequency of the normal modes in FLMS.

It may be argued that the first order temperature coefficient value of the $E^1_{2g}$ and $A_{1g}$ modes can be affected as the FLMS is supported on a $SiO_2$/Si substrate. In order to verify if the substrate has any effect on the first order temperature coefficient of the $E^1_{2g}$ and $A_{1g}$ modes, temperature dependent Raman studies on a suspended FLMS were carried out. It was observed that the phonon frequencies of both these modes red-shift with an increase in temperature and that there is no notable change in the value of $\chi$ for both the phonon modes as compared to the respective mode in supported FLMS which indicates that substrate does not have a notable effect on the temperature co-efficient.

A discussion on temperature dependent boarding of $A_{1g}$ mode is in order. Figure 4 shows the plot of FWHM of $A_{1g}$ mode as a function of temperature. Knowledge of phonon dispersion and many body theoretical calculations are essential to explain the line phonon line width of FLMS. The many body theory for $MoS_2$ crystal having several phonon branches is very complex thus we will emphasize on a qualitative description of the temperature dependence of the first order Raman scattering by considering the generalized Ridely decay channel. Accordingly we will consider that only contribution to the linewidth arises from the decay of the zone center optical phonon into one acoustic and one optical phonon. The temperature dependent phonon line width can be expressed as,[32]

$$\Gamma(T)=\Gamma_0+A[1+n(\omega_1,T)+n(\omega_2,T)] \quad (3)$$

where, $\Gamma_0$ is the background contribution, $A$ is the anharmonic coefficient, and $n(\omega, T)$ is the Bose-Einstein distribution function. Eq. 3 was fitted to the experimental data with $\omega_1=350$ cm$^{-1}$ and $\omega_2=100$ cm$^{-1}$. These two optical and acoustic phonons are chosen by considering the phonon density of state of MoS$_2$. The values of the adjustable parameters $\Gamma_0$ and $A$ are 1.8 and 0.4 cm$^{-1}$, respectively.

As discussed above both the $E^1_{2g}$ and $A_{1g}$ modes in FLMS are very sensitive to temperature change. The change in the local temperature on the FLMS due to change in incident laser power is measureable by monitoring the change in the phonon frequency. This will help us evaluating the thermal conductivity of FLMS. A supported FLMS will not be useful in this aspect as we have noticed that a change in laser power up to 2 mW does not affect the Raman peak position in supported FLMS. This is due to the faster dissipation of local heat energy to the substrate that results the Raman spectra unaffected. Thus a suspended FLMS is necessary to evaluate the true thermal conductivity. In this case the heat conduction to the sink is purely through FLMS. A typical suspended FLMS on a TEM grid (with no carbon support) is shown in Fig. 5(a). The suspended FLMS has a triangular geometry where two edges are supported on the copper grid. The length of the side edges of different suspended FLMS that we examined vary from 3 to 10 µm. The laser was focused at the centre such that the heat energy developed can uniformly propagate to the copper (heat sink). The schematic representation of laser heating on the suspended FLMS is shown in Fig. 5 (b).

In order to evaluate the thermal conductivity of the FLMS, we have made the following assumptions: (i) increase in power will not affect the temperatures of the heat sink as the laser spot on FLMS is about 1 to 1.5 µm, (ii) a steady and uniform heat conduction from the FLMS to heat sink. The expression for the heat conduction in a plane laminar surface of area $A$ can be expressed as $\frac{\partial Q}{\partial t} = -k \oint \nabla T \cdot dA$. $k$ is the thermal conductivity and $T$ is the absolute temperature. The term at the left hand side of the equation represent the rate of heat flow with time. Considering radial heat flow, Balandin et al.[19] have derived an expression for thermal conductivity of a single layer graphene as $k=(1/2\pi h)(\Delta P/\Delta T)$ where $h$ is the thickness of the layer, $\Delta P$ is the difference in laser power. As the intensity of the $A_{1g}$ mode is much stronger than $E^1_{2g}$ mode thus we consider $A_{1g}$ mode to evaluate the thermal conductivity. Previously we have discussed the linear variation (see Eq. 1) of $A_{1g}$ mode with temperature. By differentiating Eq. 1 with respect to power and substituting ($\Delta P/\Delta T$) in the above expression, the thermal conductivity can now be written as follows,

$$k = \chi_{A1g} \left(\frac{1}{2\pi h}\right) \left(\frac{\delta \omega}{\delta P}\right)^{-1}, \quad (4)$$

where $\chi_{A1g}$ is the first order temperature coefficient of $A_{1g}$ mode and $\delta\omega/\delta P$ is the change in $A_{1g}$ phonon frequency with incident laser power.

The changes in the Raman spectra of suspended FLMS with increase in incident laser power are shown in Fig.6 (a). As can be seen from the figure that the frequency of both the $E_{2g}$ and $A_{1g}$ mode red-shift systematically with increase in laser power. Secondly, these peaks have broadened significantly with increase in laser power. These changes indicate that the increase in laser power has considerably increased the local temperature on the sample surface. We have noticed in several of our samples that laser etching of the sample occurred at 0.9 mW and complete damage of the sample occurred at about 1.1 mW (**supplementary information Fig. S3**). Figure 6 (b) shows the plot of $A_{1g}$ phonon frequency as a function of incident laser power. The monotonous decrease in $A_{1g}$ phonon frequency as a function of incident laser power was best fitted with a straight line. The slope of the fitted line gives us $\delta\omega/\delta P$ and whose value is 5.7 cm$^{-1}$/mW. We have repeated this experiment several times in different samples and the value of $\delta\omega/\delta P$ is found to be 5.4±0.3 cm$^{-1}$/mW. Using the values of $\delta\omega/\delta P$, first order temperature coefficient of $A_{1g}$ mode ($\chi_{A1g}$=-1.23×10$^{-2}$ cm$^{-1}$/K) and thickness of the FLMS ($h$=6.5 nm) in Eq. 3. The extracted thermal conductivity of FLMS is found to be 52 W/mK. Recent studies of Seol *et al.*[33] on thermal conductivity of supported single layer graphene suggests that the thermal conductivity of supported single layer graphene could be suppressed significantly as compared to suspended single layer graphene (SLG). For suspended and supported SLG the thermal conductivities are found to be 5000 and 500 W/mK, respectively. The enormous suppression in thermal conductivity has been attributed to phonon leaking from the supported SLG to the substrate. As is no report on the thermal conductivity of bulk MoS$_2$ and we believe our finding will be helpful in further understanding of the thermal conductivity of supported FLMS and the effect of phonon leaking on it.

In conclusion, we report for the first time the temperature dependent Raman studies of the in-plane $E_{2g}$ and out of plane $A_{1g}$ modes in high quality FLMS prepared by vapor-phase method on SiO$_2$/Si substrate. Both the $E^1_{2g}$ and $A_{1g}$ mode red shift with increase in temperature. But, a considerable low red-shift in the frequency of the $E^2_{2g}$ mode was observed. The first order temperature coefficients of $E^1_{2g}$ and $A_{1g}$ modes are found to be $1.32\times10^{-2}$ and $1.23\times10^{-2}$ cm$^{-1}$/K, respectively. The room temperature thermal conductivity of suspended FLMS was obtained by performing a laser power dependent Raman scattering experiment on a suspended FLMS. By using the value of temperature coefficient of $A_{1g}$ mode and the data obtained from power dependent frequency shift of $A_{1g}$ mode the room temperature thermal conductivity of FLMS was extracted to be 52 W/mK.

**Acknowledgements:** The authors acknowledge partial financial supports from NSF-RII-1002410 and DOE through Grant No. DE-FG02-ER46526.

Figure Captions

Fig. 1. (a) TEM image recorded from a FLMS. (b) HRTEM image recorded from a folded edge of an 11 layers FLMS, (c, d) HRTEM images recorded from FLMS.

Fig. 2. (a) Comparison of the Raman spectra for exfoliated bulk $MoS_2$ and vapor-phase grown FLMS. (b) A schematic representation of the motion of atoms for different Raman active normal modes.

Fig. 3. (a) Raman spectra of FLMS recorded at different temperatures. (b) and (c) Temperature dependence of the frequencies of Raman active $E^1_{2g}$ and $A_{1g}$ modes in FLMS, respectively.

Fig. 4. The measured values of FWHM of $A_{1g}$ mode for the temperature in the range of 83-523K. The dashed-dot line is the fit of $\Gamma(T)$ using Eq. 3.

Fig. 5.(a) SEM image of a typical suspended FLMS (b) schematic representation of suspended FLMS, laser heating and heat conduction process.

Fig. 6. (a) Raman spectra of suspended FLMS recorded at different incident laser power. (b) laser power dependence frequency of $A_{1g}$ mode.

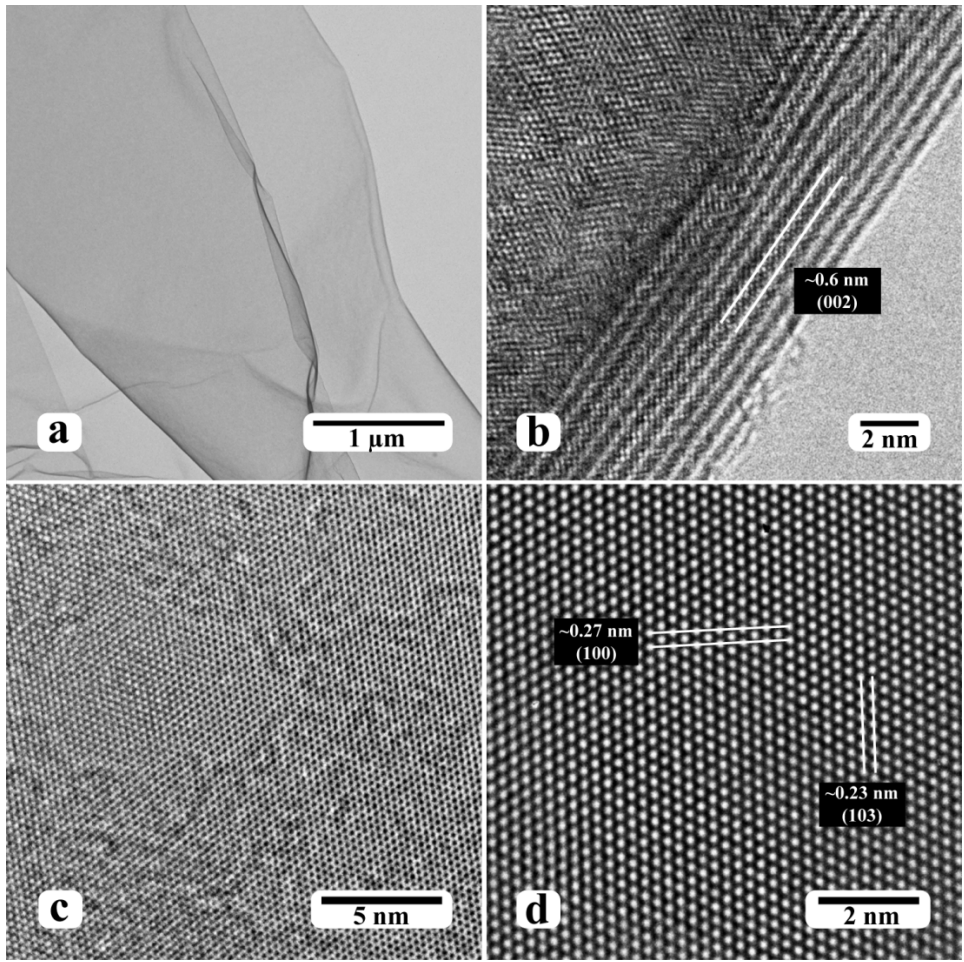

Fig. 1. Sahoo et al.

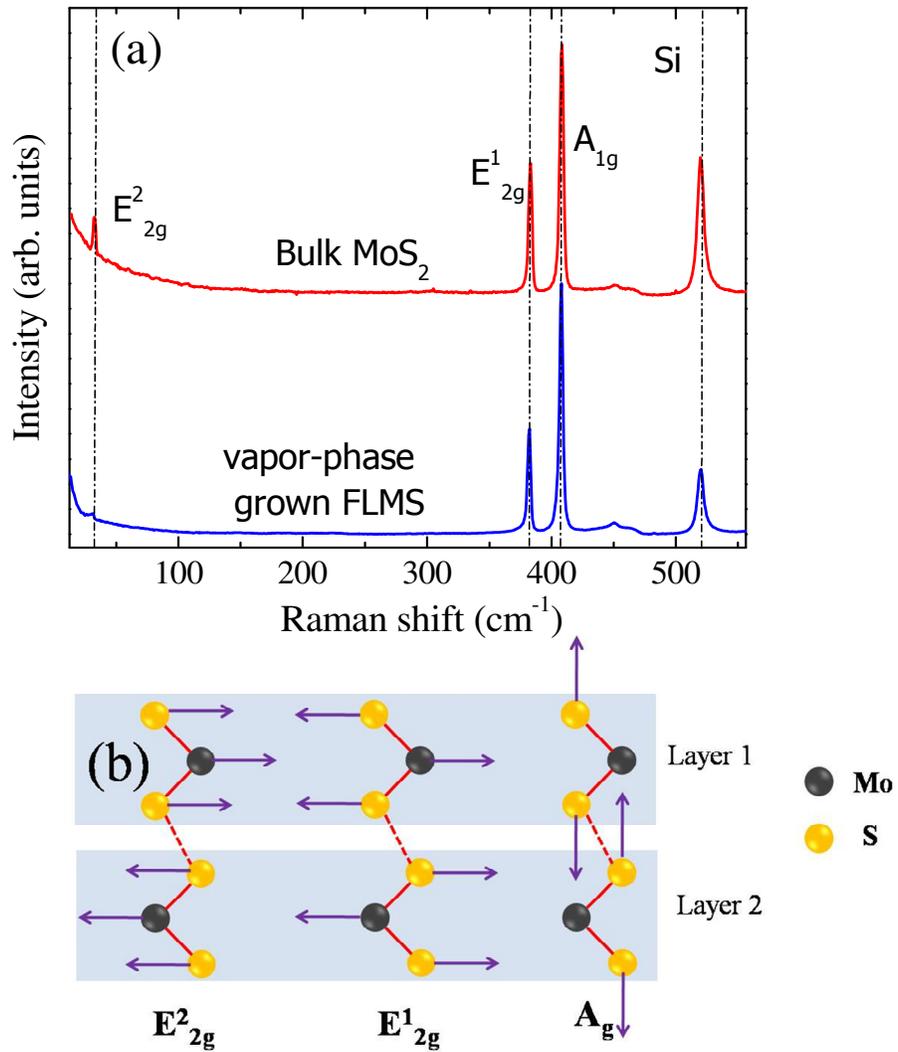

Fig. 2. Sahoo et al.

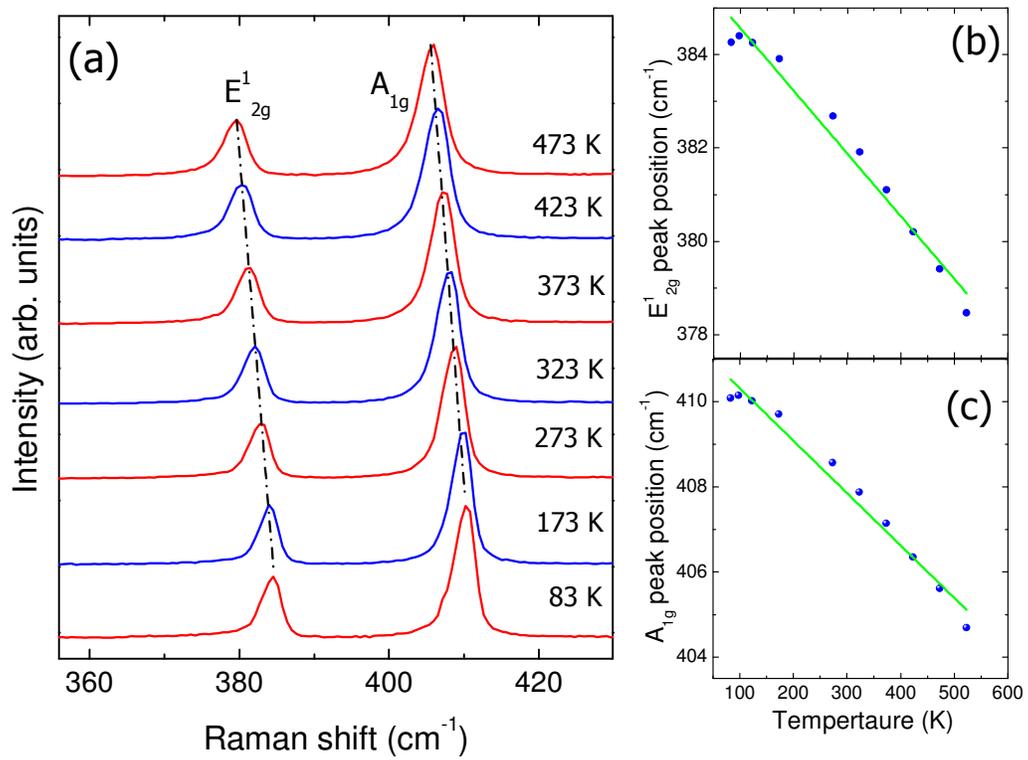

Fig. 3. Sahoo et al.

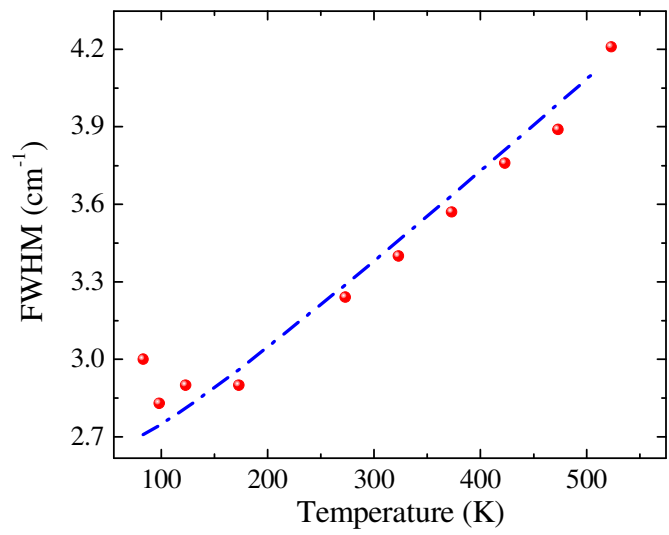

Fig. 4. Sahoo et al.

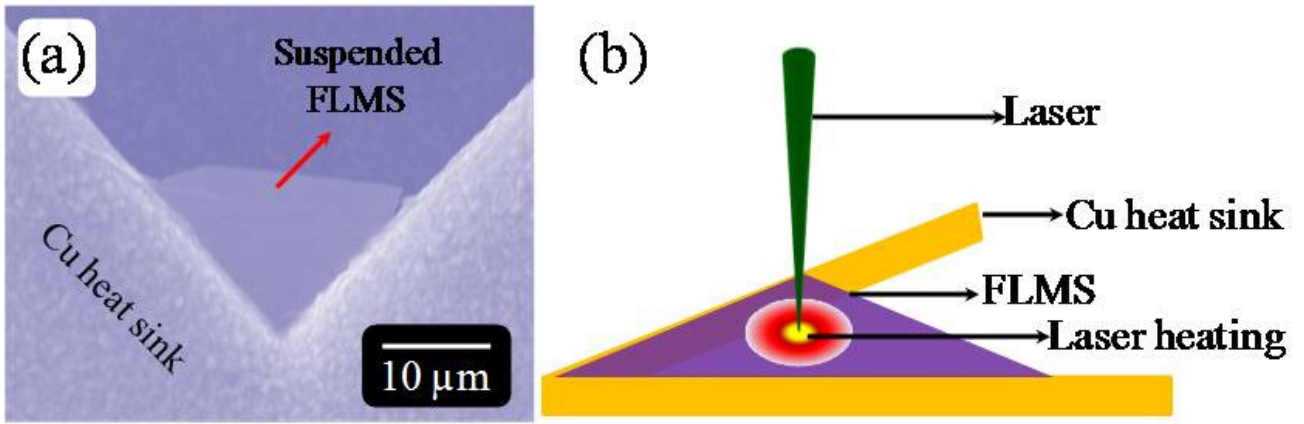

Fig. 5. Sahoo et al.

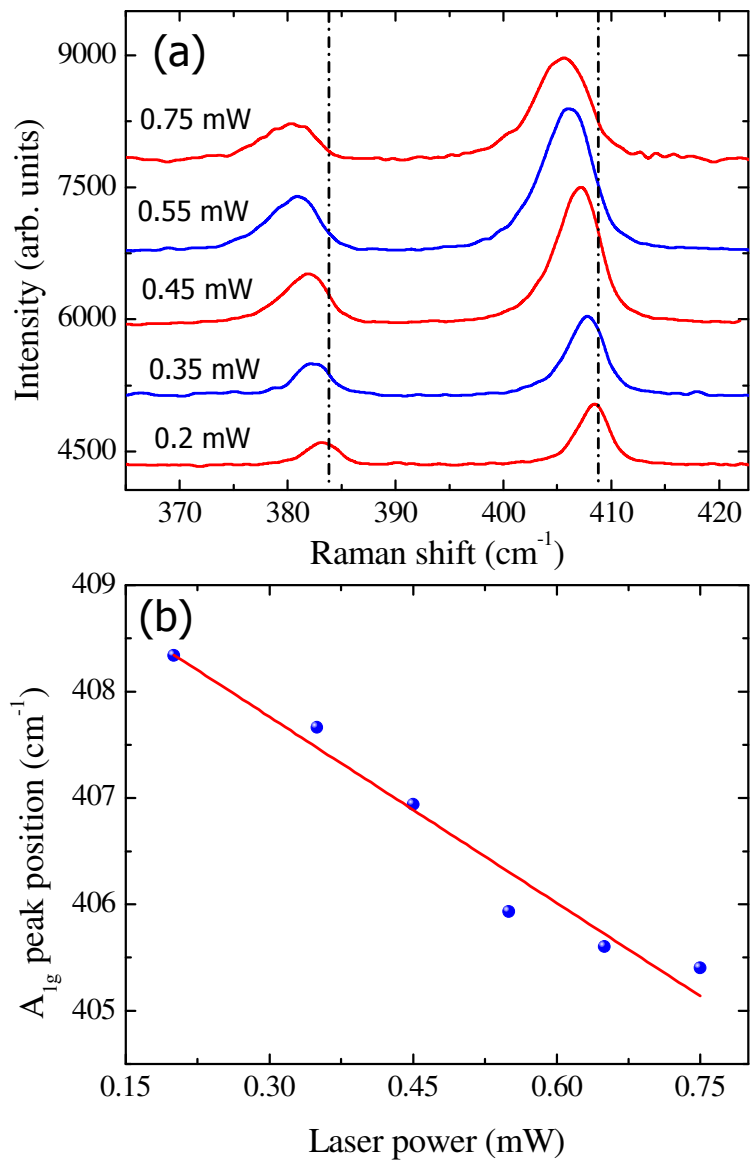

Fig. 6. Sahoo et al.

# Temperature Dependent Raman Studies and Thermal Conductivity of Few Layer MoS$_2$

Satyaprakash Sahoo,* Anand P.S. Gaur, Majid Ahmadi, Maxime J-F Guinel and Ram S. Katiyar*

Department of Physics and Institute for Functional Nanomaterials, University of Puerto Rico, San Juan, PR 00931 USA

Supplementary Information:

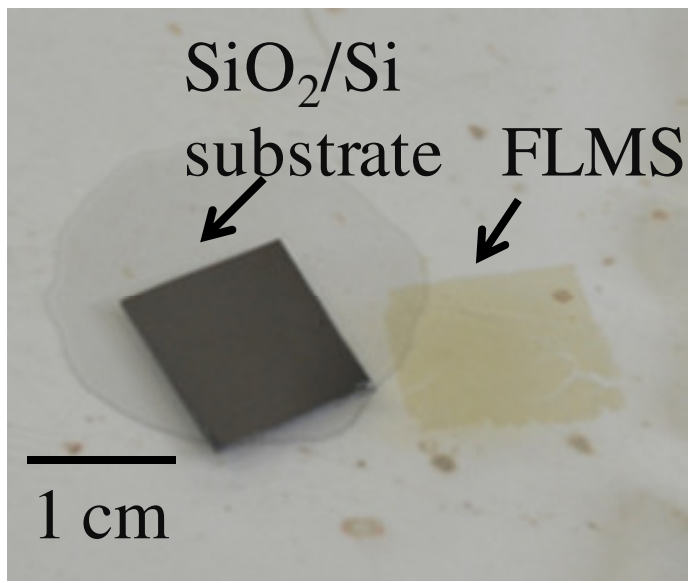

**S1. Photograph of a large scale FLMS floating on a KOH solution. It peeled off from the SiO$_2$/Si substrate.**

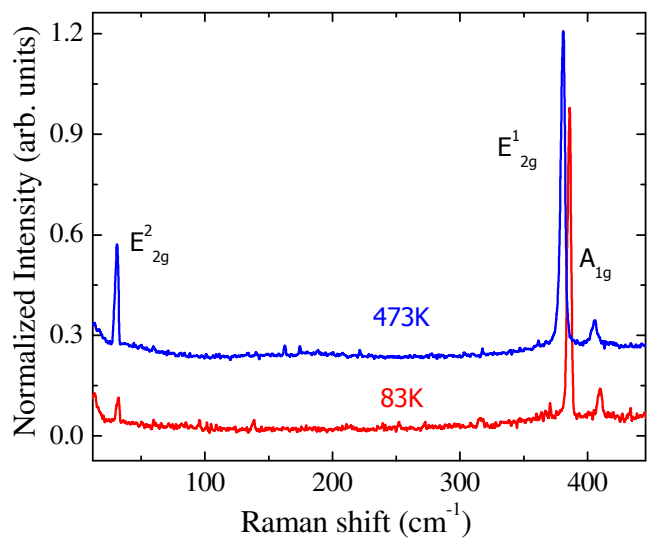

**S2. Cross polarized Raman spectra of exfoliated MoS$_2$ recorded at 83 and 473K.**

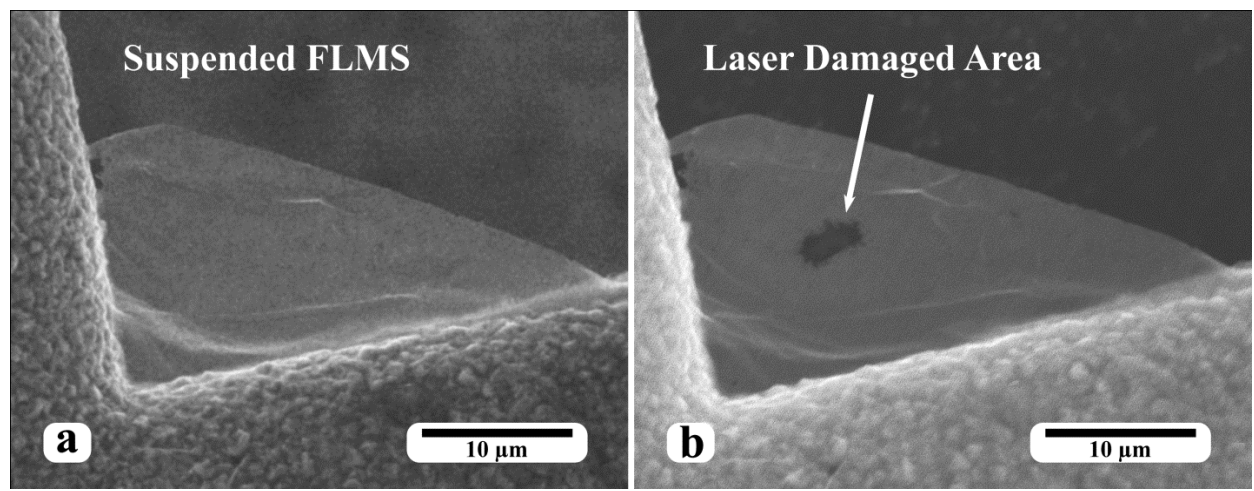

**S3. SEM image showing one suspended FLMS (a) before laser damage (b) after laser damage (the laser power was 1.1 mW).**

References:

[1] Novoselov, K. S.;  Geim, A. K.;  Morozov, S. V.;  Jiang, D.; Zhang, Y.; Dubonos, S. V.;  Grigorieva, I. V.;  Firsov, A. A. Electric Field Effect in Atomically Thin Carbon Films, *Science*, **2004**, *306*, 666-669.

[2] Zhang, Y.; Tan, Y. W.; Stormer, H. L.; Kim, P. Experimental Observation of the Quantum Hall Effect and Berry's Phase in Graphene, *Nature* **2005**, *438*, 201-204.

[3] Bolotin, K. I.; Sikes, K. J.; Jiang, Z.; Klima, M.; Fedenberg, G.; Hone, J.; Kim, P.; Stomer, H. L. Ultrahigh Electron Mobility in Suspended Graphene, *Solid State Comunn.* **2008**, *146*, 351-355.

[4] Son,Y. W.; Cohen, M. L.; Louie, S. G. Half-Metallic Graphene Nanoribbons, *Nature* **2006**, *444*, 347-349.

[5] Dean, C. R.;  Young, A. F.;  Meric, I.;  Lee, C.;  Wang, L.;  Sorgenfrei, S.;  Watanabe, K.; Taniguchi, T.;  Kim, P.;  Shepard, K. L.;  Hone, J. Boron Nitride Substrates for High-Quality Graphene Electronics, *Nature Nanotechnol.* **2010**, *5*, 722–726.

[6] Lee, K. H.; Shin, H. J.; Lee, J.; Lee I.Y.; Kim, G. H.; Choi J. Y.; Kim, S. W. Large-Scale Synthesis of High-Quality Hexagonal Boron Nitride Nanosheets for Large-Area Graphene Electronics*, Nano Lett.* **2012**, *12*,714–718.

[7] Zhu, W., Neumayer, D.; Perebeinos, V.; Avouris, P. Silicon Nitride Gate Dielectrics and Band Gap Engineering in Graphene Layers, *Nano Lett.* **2010**, *10*, 3572–3576.

[8] Kubota, Y.;  Watanabe, K.;  Tsuda,  O.;  Taniguchi, T. Deep Ultraviolet Light-Emitting Hexagonal Boron Nitride Synthesized at Atmospheric Pressure, *Science* **2007**, *317*, 932-934.

[9] Matte, H. S. S. R.; Gomathi, A.; Manna, A. K.; Datta, D. J. R.; Pati, S. K.;. Rao, C. N. R. $MoS_2$ and $WS_2$ Analogues of Graphene, *Ange.Chem.* **2010**, *122*, 4153–4156.

[10] Novoselov, K. S.; Jiang, D.; Schedin, F.; Booth, T. J.; Khotkevich, V. V.; Morozov, S. V.; Geim, A. K. Two-dimensional atomic crystals, *PNAS* **2005**, *102*, 10451-10453.

[11] Mak, K. F.; Lee, C.; Hone, J.; Shan, J.; Heinz, T. F. Atomically Thin $MoS_2$: A New Direct-Gap Semiconductor, *Phys. Rev. Lett.* **2010**, *105*, 136805-136808.

[12] Splendiani, A.; Sun, L.; Zhang, Y.; Li, T.; Kim, J.; Chim, C. Y.; Galli, G.; Wang, F. Emerging Photoluminescence in Monolayer $MoS_2$, *Nano Lett*. **2010**, *10*, 1271–1275.

[13] Cao, T.; Wang, G.; Han, W.; Ye, H.; Zhu, C.; Shi, J.; Qian, N.; Tan P.; Wang. E.; Liu, B.; Feng. Ji. Nature Commun. Valley-Selective Circular Dichroism of Monolayer Molybdenum Disulphide, **2012**, *3*, 887.

[14] Wang, H.; Yu, L.; Lee, Y. H.; Shi, Y.; Hsu, A.; Chin, M. L.; Li, L. J.; Dubey, M.; Kong, J.; Palacios, T. Integrated Circuits Based on Bilayer $MoS_2$ Transistors, *Nano Lett.* **2012**, *12*, 4674–4680.

[15] Ghatak, S.; Pal, A. N.; Ghosh, A. Nature of Electronic States in Atomically Thin $MoS_2$ Field-Effect Transistors, *ACS Nano* **2011**, *5*, 7707–7712.

[16] Wang Q. H.; Zadeh K. K.; Kis, A.; Coleman, J. N.; Strano, M. S. Electronics and Optoelectronics of Two-Dimensional Transition Metal Dichalcogenides, *Nature Nanotechnol.* **2012**, *7*, 699–712.

[17] Lee, C.; Yan, H.; Brus, L. E.; Heinz, T. F.; Hone, J.; Ryu, S. Anomalous Lattice Vibrations of Single- and Few-Layer MoS$_2$, *ACS Nano* **2010**, *4*, 2695–2700.

[18] Li, H.; Zhang Q.; Yap, C. C. R.; Tay, B. K.; Edwin, T. H. T.; Olivier, A.; Baillargeat, D. From Bulk to Monolayer MoS$_2$: Evolution of Raman Scattering, *Advan. Func. Mat.* **2012**, *22*, 1385–1390.

[19] Balandin, A. A.; Ghosh, S.; Bao, W.; Calizo, I.; Teweldebrhan, D.; Miao, F.; Lau, C. N. Superior Thermal Conductivity of Single-Layer Graphene, *Nano Lett.* **2008**, *8*, 902–907.

[20] Jo I.; Pettes, M. T.; Kim, J.; Watanabe, K.; Taniguchi, T.; Yao, Zhen.; Shi, L. Thermal Conductivity and Phonon Transport in Suspended Few-Layer Hexagonal Boron Nitride, *Nano Lett.* 2013, *13*, 550-554.

[21] Slack, G. A. Anisotropic Thermal Conductivity of Pyrolytic Graphite, *Phys. Rev.* **1962**, *127*, 694-701.

[22] Zhan, Y.; Liu, Z.; Najmaei, S.; Ajayan, P. M.; Lou J. Large-Area Vapor-Phase Growth and Characterization of MoS2 Atomic Layers on a SiO2 Substrate, *Small* **2012**, *8*, 966-971.

[23] Lee, Y. H.; Zhang , X. Q.; Zhang, W.; Chang, M.T.; Lin , C. T.; Chang, K. D.; Yu, Y. C.; Wang, J. T.W.; Chang, C. S.; Li, L. J.; Lin, T. W. Synthesis of Large-Area MoS$_2$ Atomic Layers with Chemical Vapor Deposition, *Adv. Mater.* **2012**, *24*, 2320–2325.

[24] Sahoo, S.; Arora, A. K. Laser-Power-Induced Multiphonon Resonant Raman Scattering in Laser-Heated CdS Nanocrystal, *J. Phys. Chem. B* **2010**, *114*, 4199-4203.

[25] Gupta, R.; Xiong, Q.; Adu, C. K.; Kim, U. J.; Eklund, P. C. Laser-Induced Fano Resonance Scattering in Silicon Nanowires, *Nano Lett.* **2003**, *3*, 627.

[26] Ataca, C.; Topsakal, M.; Akturk, E.; Ciraci, S. A Comparative Study of Lattice Dynamics of Three- and Two-Dimensional $MoS_2$, *J. Phys. Chem. C* **2011**, *115*, 16354-16361.

[27] Plechinger, G.; Heydrich, S.; Eroms, J.; Weiss, D.; Schüller, C.; Korn, T. Raman Spectroscopy of the Interlayer Shear Mode in Few-Layer $MoS_2$ Flakes, *Appl. Phys. Lett.* **2012**, *101*, 101906-101908.

[28] Loudon, R. The Raman Effect in Crystals, *Advanc. in Phys.* 1964, *13*, 423-482.

[29] Calizo, I.; Balandin, A. A.; Bao, W.; Miao F.; Lau, C. N. Temperature Dependence of the Raman Spectra of Graphene and Graphene Multilayers, *Nano Lett.* **2007**, *7*, 2645-2649.

[30] Sahoo S.; Arora, A. K.; Sridharan, V. Raman Line Shapes of Optical Phonons of Different Symmetries in Anatase $TiO_2$ Nanocrystals, *J. Phys. Chem. C* **2009**, *113*, 16927.

[31] Peercy, P. S.; and Morosin, B. Pressure and Temperature Dependences of the Raman-Active Phonons in $SnO_2$, *Phys. Rev. B* 1973, *7*, 2779–2786.

[32] Menéndez, J.; and Cardona, M. Temperature Dependence of the First-Order Raman Scattering by Phonons in Si, Ge, and *α*-Sn: Anharmonic Effects, *Phys. Rev. B* 1984, *29*, 2051.

[33] Seol, J. H.; Jo, I.; Moore, A. L.; Lindsay, L.; Aitken, Z. H.; Pettes, M. T.; Li, X.; Yao, Z.; Huang, R.; Broido, D.; Mingo.; Ruoff, R.; Shi, L. Two-Dimensional Phonon Transport in Supported Graphene, *Science* **2010**, *328*, 213-216.